\documentclass[conference]{IEEEtran}

\usepackage{ifpdf}
\usepackage{cite}

\ifCLASSINFOpdf
  \usepackage[pdftex]{graphicx}
  \DeclareGraphicsExtensions{.pdf,.jpeg,.png}
\else
  \usepackage[dvips]{graphicx}
  \DeclareGraphicsExtensions{.eps}
\fi

\usepackage[cmex10]{amsmath}

\usepackage{fixltx2e}
\usepackage{url}
\usepackage{algorithm}
\usepackage{algorithmic}
\usepackage[caption=false,font=footnotesize]{subfig}

\begin{document}

\title{Toward interoperability for the Internet of Things with meta-hubs}

\author{\IEEEauthorblockN{Julien Mineraud and Sasu Tarkoma}
\IEEEauthorblockA{Department of Computer Science\\
University of Helsinki\\
Finland\\
Email: fistname.lastname@cs.helsinki.fi}}

\maketitle

\makeatletter{}\begin{abstract}
The Internet of Things (IoT) envisions that objects may be connected to the Internet, producing and consuming data in real-time.
Today, numerous middleware platforms are available to facilitate the communication with these objects. 
Unfortunately, the interoperability of these platforms is very limited because it requires to ``manually'' connect the services proposed by each platform. 

One key design goal for our contribution is not to build yet another middleware, but rather to augment the functionalities of existing systems via an extension
to support their integration into a network of heterogeneous IoT hubs.
The extension includes a RESTful API to manipulate the basic component of our extension, the IoT feeds. 
The IoT feeds allow the platform's owner to dynamically marshal the IoT features connected to the platform, as well as the data that they produce.
Furthermore, the feeds enable the owner to manage and control the data flows before connecting them to his applications.
Subsequently, these feeds may also be published to meta-hubs in order to expose them to third parties. 

We evaluated an implementation our extension for Android systems to show the feasibility of managing the data flows using the RESTful API on this platform.

\end{abstract} 

\IEEEpeerreviewmaketitle

\makeatletter{}\section{Introduction}\label{sec:introduction}

The Internet of Things (IoT) envisions to combine ``smart'' objects in our environment into a fully integrated future Internet~\cite{Atzori2010}.
Today,  plethora of IoT middleware systems, proprietary or open-source, are available for integrating these smart objects into the IoT.
The key priority for these platforms is usually to establish communications with a large variety
of smart objects, such as sensors and actuators. These platforms also have varying requirements and technological implementations
depending on the type of uses of the platform (e.g. domotic, environment monitoring, production line control, etc\ldots). 
Furthermore, these platforms generally have their own ecosystems and may provide tools for applications or services developers (e.g. libraries, SDK, etc.). 

However, IoT systems would benefit from using a generic architecture which provides IoT developers with a common toolkit for the development of IoT services 
independently of the underlying systems.
We selected the following key principles for this architecture:
\begin{itemize}
 \item Interoperability with other systems
 \item Ownership of data and device functionalities
 \item Scoping for managing data and functionality visibility
 \item Loose coupling and late binding
\end{itemize}
The first principle is the interoperability of the platform with other systems in order to form networks of heterogeneous but interoperable IoT systems. Moreover,
the second principle ensures that the platform user retains full control of his devices and the data that they produce. This would be characterized by private and global spaces.
Within the private space, resources are accessed via state-of-the-art access control mechanisms (e.g. OAuth2, etc.), while in the global space, resources can be accessed from
within the network of interconnected platforms. This leads to the third principle where scopes are defined to grant permissions to use/view the data and other functionalities
that are provided by the platform. Finally, the loose coupling and late binding designs have been selected to ease the manipulation of IoT resources by IoT developers using the
common toolkit, in both private and global spaces.

In this paper, we present the IoT Hub architecture that covers these essential characteristics.
Our solution is a simple extension of available middleware systems to provide communication between the middleware system and the IoT, as well as tools to control the data flows. 
The extension includes a RESTful API that supports fine-grained data management while re-attributing the ownership of data and devices to the hub owner.
The IoT Hub API defines the concept of IoT feed (loose coupling) which is the basic component that marshals IoT data and features (i.e. sensors, actuators) available within the platform.
An IoT feed is composed of its metadata (e.g. time-serie data, dependencies to other feeds, scopes, keywords, etc\ldots) as well as the declaration
of fields metadata (e.g. type, name, keywords, etc.), and if applicable its data. 
The fields are not bound by numbers, but potentially by their types. In fact, the IoT feed metadata declaration ensures that fields have compatible types. 
For instance, a field that access a sensor's data cannot be included in an IoT feed describing time-series data. 
The IoT feeds are exposed locally via the IoT Hub API for composition within IoT services (late binding), reserving that the IoT services have sufficient permission to access the feeds
(fine-grained data flow control).
Moreover, the hub owner has the option of publishing IoT feed descriptions to a special type of IoT hub, the meta-hub. 
Meta-hubs are the bridging pieces toward interoperability for hub-based middleware systems.
The main role of the meta-hub is not to connect smart objects to the platform, but to collect IoT feeds metadata that are published by IoT hubs into a browsable catalog.
Additionally, meta-hubs include catalogs for IoT applications and services, as well as a search engine to extract information from their catalogs. 

In order to execute the services and applications provided by the meta-hubs, the hub architecture includes a javascript engine. The choice of javascript was driven
by the fact that most current IoT systems provides web services (RESTful API) and applications are included into a dashboard that is accessed via a web browser. 
Therefore, these IoT platforms could implement the IoT Hub architecture with only a small overhead.

\noindent The contributions of this paper are the following:
\begin{itemize}
 \item We present a simple yet powerful IoT platform concept, the IoT hub, that interconnects various existing IoT systems. 
 The main novelty stems from the IoT Hub API that exposes both data and control features to third parties.
 \item The hub follows a data feed design that allows fine-grained and real-time management of flows within a hub and between hubs (i.e. private and global (public) spaces).
 \item We present a concrete toolchain and implementation of the IoT hub for Android systems and demonstrate the benefits of local data feed processing.
\end{itemize}

This paper is organized as follows. Section~\ref{sec:relatedWork} provides an overview of IoT middleware platforms and solutions toward interoperability for the IoT.
This will be followed by Sections~\ref{sec:iothub} and~\ref{sec:metahubs} presenting respectively the IoT hub and the meta-hubs.
Section~\ref{sec:ahub} presents our implementation of IoT hub for Android devices and its evaluation will be presented in Section~\ref{sec:evaluation}.
Finally, Section~\ref{sec:conclusion} concludes this paper and presents the directions for our future work.
 
\makeatletter{}\section{Related work}\label{sec:relatedWork}

In this section, we will present a number of IoT middleware solutions that could to be extended as IoT hubs in the future. 
This will be followed be a summary of efforts that have been made toward the interoperability for the IoT.

\subsection{Today's IoT platform landscape} 

There are numerous IoT platform available today. The following list is not exhaustive but summarizes well the current landscape of IoT solutions.

The EveryAware platform~\cite{Becker2013} provides an extendable data concept that could be use to enhance the 
possibilities of sharing and fusing data feeds. The platform is a centralized solution and was the only one focusing on a fine-granularity of data
visibility. In our opinion, fine-grained data visibility is essential but is not yet well covered by current IoT solutions. 
In addition, it provides flexible and extendable data models.

The LinkSmart\footnote{\url{http://www.hydramiddleware.eu/news.php}} middleware platform,
formerly Hydra, is an open-source service-oriented platform that enables the creation of a network for embedded systems, using semantics to discover the devices connected to the network.

The OpenIoT\footnote{\url{http://openiot.eu/}} platform is an open-source platform, fully decentralized, 
that provides connectivity with constrained devices such as sensors. 
The platform envisions a billing mechanism for the use of services that would integrated to their solution.

The Thing System\footnote{\url{http://thethingsystem.com/}} is a open-source software using Node.js that enables the discovery of smart things in the home environment.
The software does not provide storage functionalities and must be coupled with a PaaS (i.e. Platform-as-as-Service) to enable storage outside the home area. 

ThingSpeak\footnote{\url{https://www.thingspeak.com/}} is a decentralized, open-source server that may be used to store and retrieve IoT data. 
It allows opening of the channels to the public but do not provide extensive configuration of the these channels. 
The platform also provides visualization tools and enables the creation of widgets to visualize the data in a more personified fashion.

The Thing Broker~\cite{PerezdeAlmeida2013} is a centralized platform that provides a Twitter-like abstraction model for things and events to create local ecosystems such as smart homes.
All the platforms previously mentioned, apart from OpenIoT, provide a RESTful API to interact with the platform.

\subsection{Toward the interorability for the Internet of Things}

Recent work on distributed services for the IoT have shown the tremendous potential of supporting interoperability amongst IoT devices, platforms and users.
For instance, Rachuri et al.~\cite{Rachuri2013} proposed a distributed task allocation mechanism from the phone to the 
local infrastructure to save battery energy of the mobile devices and thus, make the sensing as less intrusive as possible for the end users. 
Their solution achieved, by offloading sensing task to the local infrastructure, an energy saving of 35\% compared to pure sensing.
The offloading of sensing depending on the context-awareness have also been achieved with CoSense~\cite{Hemminki2013} that enable sharing of GPS location 
for users using a common transportation (e.g. in a train, bus, tramway or car). 

However, a number of challenges, such as service composition, data-points availability, service discovery or data accuracy,
have been highlighted in~\cite{Teixeira2011,Cherrier2013, Xiao2014} to create interoperable service-oriented IoT solutions.
Unfortunately, these challenges have not been addressed by today's IoT solutions.

Lea and Blackstock~\cite{Blackstock2013,Lea2014} have recently studied of the interorability of IoT platforms. Resulting from the observation of numerous IoT projects
within the United Kingdom where most solutions are not interoperable, the authors presented a set of solutions to address this challenge. In particular, the authors introduced
a multi-stage solution for interoperable IoT deployments~\cite{Blackstock2013}. 

The first stage consists on enabling IoT platforms to expose their services via RESTful APIs.
Currently, most existing IoT platforms provide a RESTful API to enable interactions between the user of the platform and the smart objects, as well as extracting time-series data
from the storage. At this point, IoT service developers can use these services at a high cost of ``manually'' connecting the services to their applications.

The second stage involves agreements of commun understanding, such as abstracted data and service models, between interoperable actors. The authors introduced for this latter stage,
the \emph{HyperCat} protocol~\cite{Lea2014}, which aims to provide generic modeling of IoT data and services in order to construct catalogs of IoT services (URI) that are proposed by IoT
platforms publicly.

The last stage includes the integration of complex semantics and ontologies to describe the services and things accessible via these platforms. We believe that such staged development
for interoperability between IoT platforms, things and services is necessary to take advantage of the possibilities offered by a fully connected future Internet. 

However, the catalog server do not incorporate billing and accounting mechanisms to facilitate transactions between IoT hubs, and the design of the search engine is currently based only on filtering. 
In the next section, we will present a simple yet powerful IoT platform concept that includes the basic functionalities to cope with the challenges inherited from the foreseen 
properties of the IoT.

%XMPP to bridge the gap of Internet of things~\cite{Kirsche2012} 
\makeatletter{}\section{IoT Hub}\label{sec:iothub}

Based on our experience with IoT middleware solutions, the heterogeneity of these platforms is considerable (numerous hardware, technologies,
requirements and objectives).
Unfortunately, current platforms are not able to provide a solution that adapts to the requirements and objectives of all users. 
For instance on the first hand, The Thing System, which was mentioned in Section~\ref{sec:relatedWork}, is designed for home automation
and aims to simplify the communication between the system owner and his sensors and actuators (e.g. switch on or off a smart light,
reading room temperature, etc.) and offers a nice and easy-to-use interface to the end-users. 
On the other hand, platforms, such as Axeda's\footnote{\url{http://www.axeda.com/}}, provides PaaS for
machine-to-machine (m2m) interactions to businesses in order to perform analytics and make proactive actions to replace faulty assets or 
schedule the recalibration of sensors.

\begin{figure}[!t]
 \centering
 \includegraphics[width=.65\linewidth]{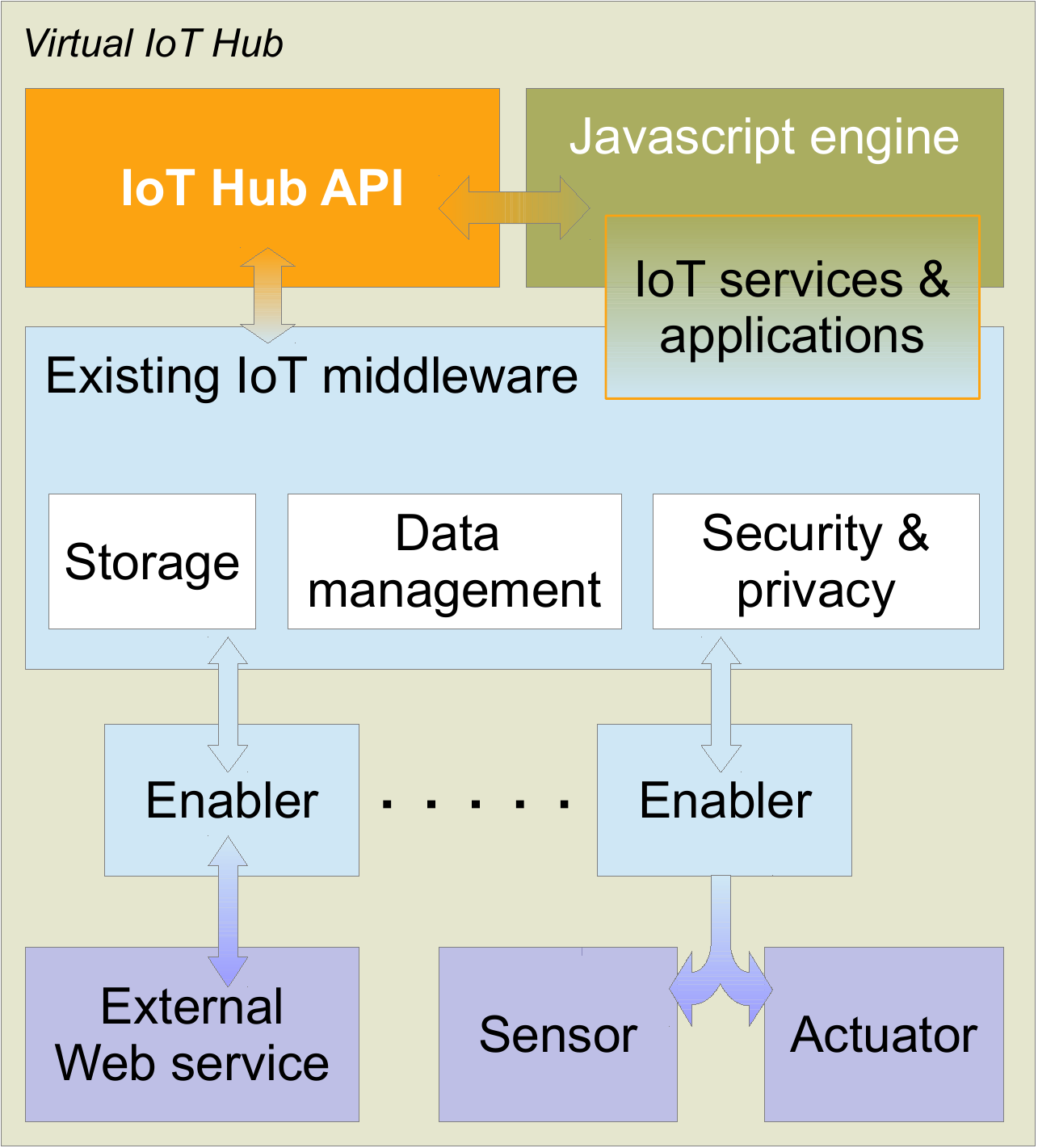}
 \caption{IoT Hub architecture}
 \label{fig:architectureHub}
\end{figure}

Therefore, as our main objective is to create a network of heterogeneous hubs (depicted in \figurename{~\ref{fig:metaHubs}}), 
we propose the IoT Hub architecture (see \figurename{~\ref{fig:architectureHub}}). 
It has been designed to be easily implemented on any exisiting middleware systems.
Hence, the architecture is independent of underlying technologies such as database management systems or particular web server implementations.

The IoT hub architecture is generic and includes a common stack of protocols to handle the devices or the data that
would be processed by the hub. The hub may be small, if owned by a single individual (e.g. household), or very large
(e.g. major company).
The hub relies on the existing middleware system to enable the communication with the IoT features and external web service.
The underlying system is also responsible of the storage of data~\cite{Roman2011}, as well as the privacy and security mechanisms to access the platform~\cite{Ma2013,Wang2014}.

Depending on the hardware running the hub and the objectives of the IoT solution, several technologies could be used to
fulfill these tasks. For example, IoT hubs running on mobile devices are limited in storage
capacity, require complex energy management policies and are bound by the operating system to use specific technologies.
From another standpoint, an instance of cloud-based IoT hub may have significantly more flexibility.
In the particular case of data storage, some of the existing IoT platforms rely on NoSQL databases, such as \emph{MongoDB} for the
storage of time-series of sensor data, while others rely on large SQL databases such as PostgreSQL or Oracle. 

The IoT hub is dependent on the underlying platform to provide enablers to instantiate the communication between the smart objects and the
RESTful API. In the case of The Thing System, more than 70 ``things'' are currently supported by the platform, and 30 more are in
development stage. We desire that IoT hub implementations beneficiate from precedent efforts on enabling IoT technologies. 
The enablers also includes bridges to external web services (e.g. google maps API) in order to enrich the local data with external information.

In the IoT hub architecture, these enablers are also responsible of transforming the things to IoT feeds complying to the IoT Hub API. 
Our architecture favors the loose coupling of things and data by manipulating them via the concept of IoT feed. 
The IoT feed is the basic component of our architecture to abstract and marshal the properties of the things or the data that they produce.
IoT feeds also include a non-empty set of fields that is not bound by size, but potentially by their types to ensure consistency of the fields
with respect to the feed properties.
As an example, a time-series feed, which may be composed or several data types (e.g. timestamp + temperature + GPS location), does not have the same properties 
than the atomic feed that accesses the current reading of a sensor or the state of an actuator.

From the IoT hub viewpoint, the things and the data that are handled by the platform are all represented by IoT feeds, which can be manipulated and combined within
the private local space of the platform. 
To be noted that IoT hub feeds declarations are dynamical. A publish/subscribe mechanism is also available to interconnect feeds and generate data flows.

\begin{figure}[!t]
 \centering
 \includegraphics[width=.65\linewidth]{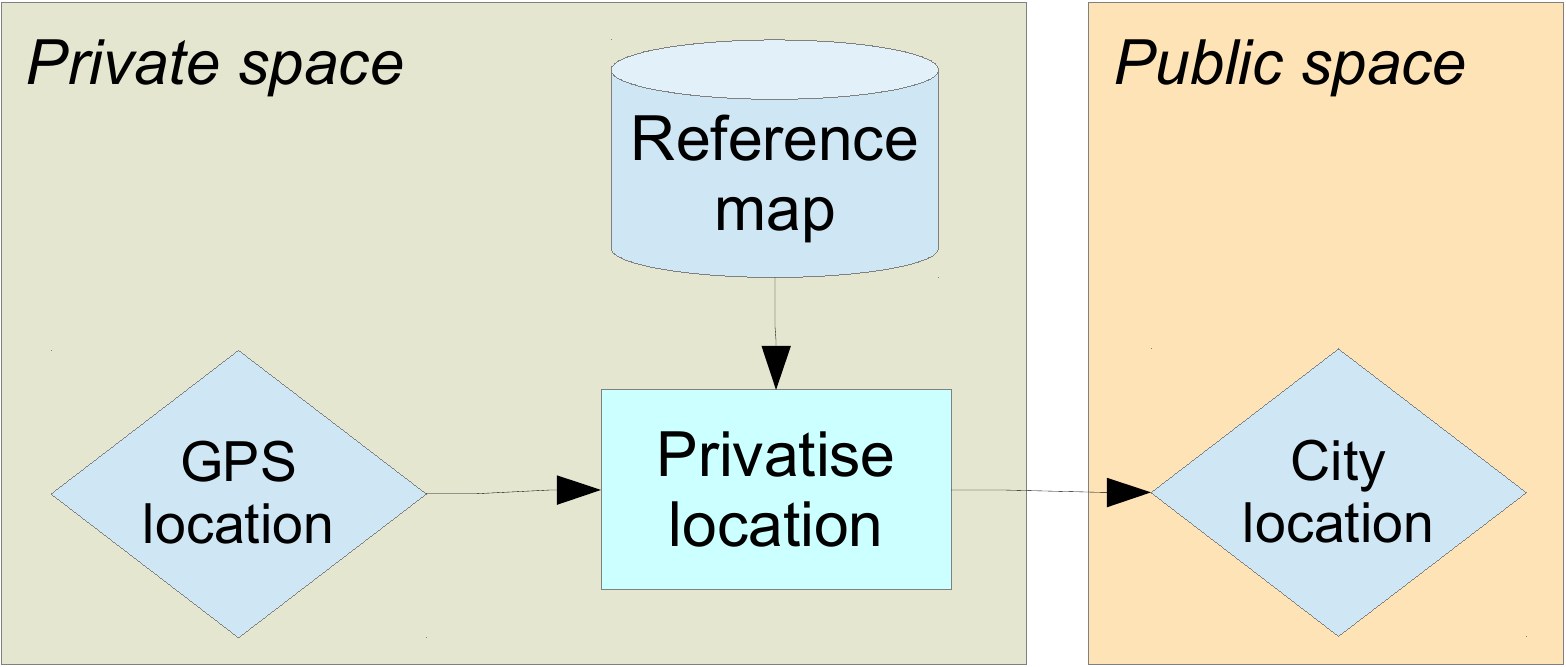}
 \caption{Example of IoT hub feed manipulation}
 \label{fig:dataFlow}
\end{figure}

The local management of feeds will provide a rich environment for the IoT hub owner to manipulate data with increased flexibility. 
Consequently, the IoT hub architecture requires a strong modeling of data and services interoperability with strong type checking (data representation) and 
well-defined operations that can be applied to feeds (e.g. temperature data can not be aggregated with relative humidity data although they can be both represented by decimals).
We envision for the back-end system of an IoT hub, a similar approach to the \emph{Yahoo! pipes\textsuperscript\texttrademark} service that transforms RSS feeds (from multiple sources)
into new composed RSS feeds. In \emph{Yahoo! pipes\textsuperscript\texttrademark}, simple operations such as filtering or text search
can be applied to RSS feeds to enrich the content of the newly generated feed. Transposed to our IoT approach, IoT feeds could be derivated locally from local and external
sources of information to be exposed to third-party services via the RESTful IoT Hub API.
Furthermore, IoT services deployed on the platform could trigger operations based on the data they monitor, in a similar fashion than the IFTTT platform\footnote{\url{https://ifttt.com/}},
or modify the sampling rate of time-series data to either increase accuracy or reduce the need of storage space.  

A simplified application of this process with respect to the privacy of the user would be the transformation of the exact location of a hub user (GPS coordinates) into an 
anonymised location feed that only tells in which city the user is presently. 
This new feed could then be shared to external services that require GPS location only at a city-level granularity.
An example for this third-party service would be a hub owned by a city and collecting weather measurements of individuals currently located in the city. 
The data flow of this example is shown in Figure~\ref{fig:dataFlow}.
Consequently, new fine-grained policies for data exchange could be implemented and give the hub owner full control on the type of information that
is shared externally (accessible from the public space). 

The last feature of the IoT Hub architecture is an application engine to execute IoT applications that are based on the IoT Hub API. 
We opted for a javascript engine because most of today's IoT applications are run from a dashboard in a web browser. Hence the
majority of today's IoT platforms would easily be extendable with no overhead.

In the next section, we will describe a particular type of hub, called meta-hub which provides additional functionalities to support interoperability
between heterogeneous hub platforms.

\section{Meta-hubs for IoT interoperability}\label{sec:metahubs}

\begin{figure}[!t]
 \centering
 \includegraphics[width=.65\linewidth]{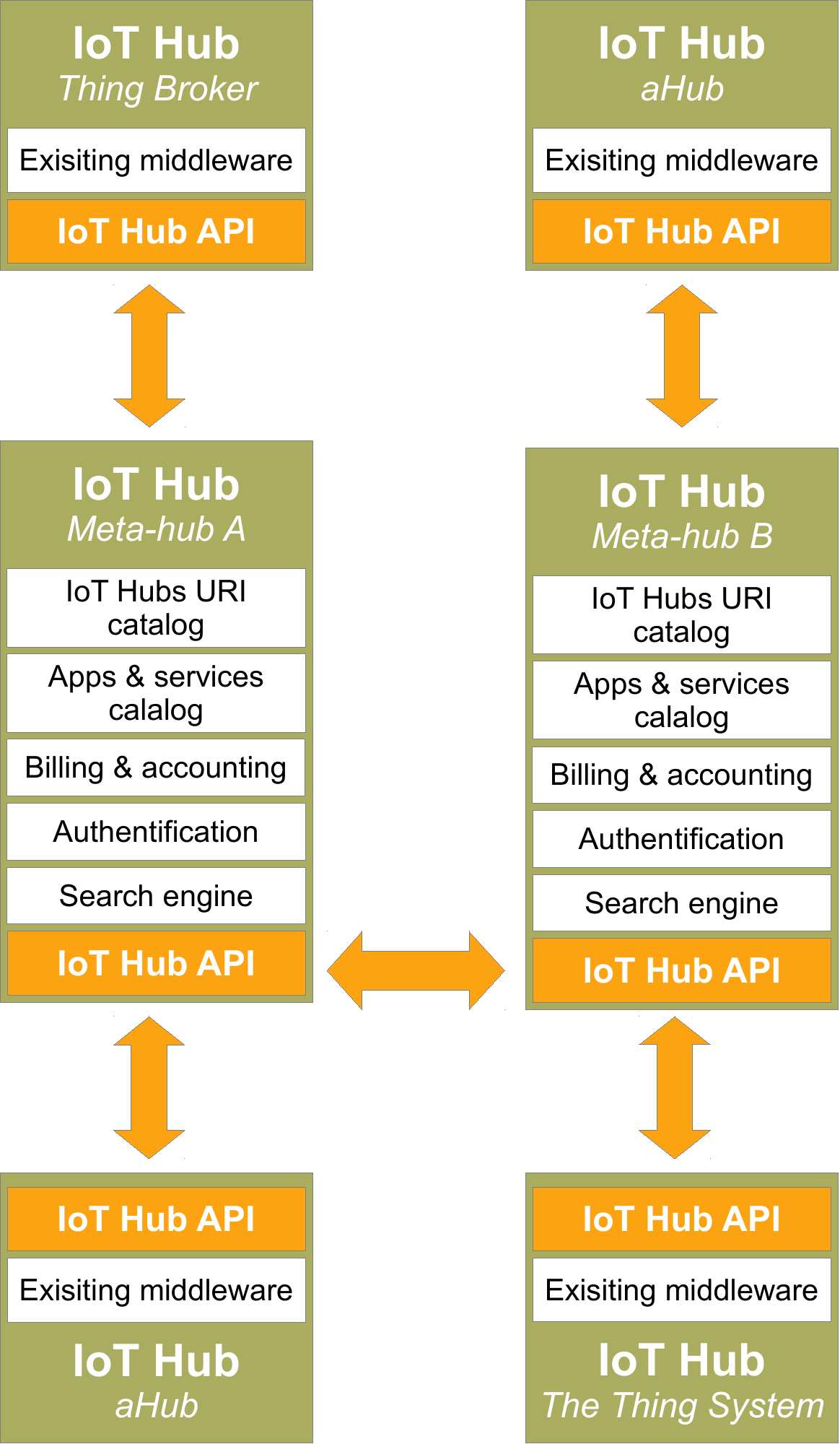}
 \caption{Network of IoT hubs including meta-hubs}
 \label{fig:metaHubs}
\end{figure}

Figure~\ref{fig:metaHubs} depicts the internal components of meta-hubs. Meta-hubs are a variation of regular IoT hubs with a few
differences. The role of meta-hubs is not to enable communication between the platform and smart objects, even if this is not mandatory, 
but rather to store information about IoT hubs and the services that they published. 
We opted for an IoT Hub URI catalog that is similar to \emph{HyperCat}~\cite{Lea2014} which is a repertoire of IoT hub URI that were published by IoT hubs. 

Additionally, the meta-hubs would include another catalog for applications and services that are based on the IoT Hub API and can be executed by the
javascript engine. This allows the IoT developers to publish and distribute their products to a large number of IoT platforms effortlessly.
Moreover, the applications can take advantage of using local feeds while extracting data from remote sources that are available through the meta-hubs.
For example, an application could extract temperature from a local sensor feed, and compare the recorded values with the twenty closest temperature sensor
feeds from the current location of the user without needing to manually set with which IoT platforms the application has to connect.

To realize these operations, meta-hubs need to incorporate an efficient search engine to browse efficiently through their own IoT feed catalogs, as well as requesting
missing information from other meta-hubs. Recent work~\cite{Ostermaier2010} investigated the exploitation of the current web infrastructure to build a search engine dedicated
to the IoT. However, we believe that the catalogs of meta-hubs will bridge that gap and the main attributes of the search engine would be to retrieve the best quality of results 
(e.g. data accuracy, low latency) while avoiding data redundancy.

In addition, components for authentication, billing and accounting could be included to meta-hubs to facilitate transactions between IoT hubs 
and create potentially novel business and economical models taking advantage of the new opportunities emerging from the IoT. 
An example is the Windows Azure Data Market\footnote{\url{http://datamarket.azure.com/}} which provides catalogs of data which are browsable online, so application developers can 
easily find the resources necessary to make their application. 
The marketplace includes different billing schemes (e.g. time-based, quantity-based, free, etc.). The disadvantage of their solution is the need for data producers to build models 
for data manipulation which require the application developers to learn the data models for every dataset.

Lastly, we would like to elaborate on a possible application and show where the meta-hubs and IoT hub infrastructure would improve existing infrastructures. 
In a smart city scenario, we foresee that IoT hubs could be owned by individuals in every home. The IoT hub would provide off-the-shelf applications for home automation 
(i.e. retrieved from the meta-hubs application catalog).
Today, home automation middleware systems enable the users to interact with their devices wirelessly, monitor various environmental parameters such as the electric
consumption, room temperature and even the content of their fridge. Mostly this information is private, but home owner may be willing to send weekly
reports (privatized feeds) about their energy consumption to the building hub. In the meantime, the building hub may build its own weekly
report for the city hub, which may be able to analyze the data and deduct if a particular neighborhood would have abnormal energy consumption. 
The city may in the future understand that this neighborhood may, for instance, have a different isolation material that is inefficient for this location and act accordingly.

Consequently, to achieve this type of scenario, a large dissemination of heterogeneous IoT hubs and meta-hubs would be necessary. 
Hence, we present in the next section an implementation of the IoT hub for Android systems.

\section{IoT Hub for Android}\label{sec:ahub}

We implemented a version of the IoT hub for Android (\emph{aHub}) because the majority of IoT platforms for mobile phones are rather limited and that mobile phone users represents
a potentially large volume of IoT users for the IoT hub architecture to sustain (i.e. ecosystems of IoT hubs). 

The IoT hub for Android uses the NanoHttpd\footnote{\url{http://nanohttpd.com/}} lightweight web server to answer HTTP requests as 
defined by the RESTful IoT Hub API. The application includes two activities: (i) a WebView that loads a local HTML webpage where
javascript applications are available. An example of this application will be presented in Section~\ref{sec:evaluation}.
The second activity provides an interface to transform the features of the Android devices info IoT feeds.
This activity lists all the available enablers on the platform. If an enabler is missing (e.g. exotic device that was just purchased), 
the user could download from the meta-hub the desired enabler to its hub and use it on the fly.

A key design goal of the IoT hub architecture is that the IoT hub provide full control to the users. As a result, we designed the original IoT hub for Android phone
to require no permission during installation. The features that requires additional permissions would need to be downloaded from the meta-hubs and installed as in-apps.

\subsection{Toolkit for IoT services development}

We propose the javascript language to develop IoT applications and services for the IoT hub because of the inherited web nature of the IoT hub.
We use the \emph{js\_of\_ocaml}~\cite{Vouillon2013} compiler to develop IoT applications for IoT hubs. The original application is developed in the Ocaml
language. Ocaml is a functional programming language with very interesting features to build applications for our proposed platform.

First, it is characterized by its strong static data typing and second, its efficiency of manipulate complex data structure. 
As one of our design goals is to incorporate to the IoT Hub API a complex model of relations between IoT data types and services, we used Ocaml to design reliable and efficient IoT services. 
Moreover, Ocaml has numerous available modules to facilitate the development of Ocaml applications (e.g. Lwt for cooperative threading which we have been using extensively to
handle requests to IoT hub web server). 
Finally, the Ocaml syntax can be extended with limited overhead for the development of a Domain Specific Language (DSL) dedicated to the IoT. 
The choice of Ocaml was driven by the prospect of using an IoT DSL based on the IoT Hub API for the development of our future applications.
However, we did not include the DSL implementation to the scope of this paper.

In the next section, we will present the evaluation of the IoT hub implementation for Android devices.
 
\makeatletter{}\section{Evaluation}\label{sec:evaluation}

To evaluate the IoT hub platform and demonstrate the feasibility of managing the data flows using the IoT Hub API, we designed a simple application. 
The application first checks for the availability of two types of atomic feeds: (i) a feed giving access to accelerometer data and 
(ii) a feed to a \texttt{ON/OFF} switch (data modeling of an actuator with only \texttt{ON} and \texttt{OFF} states).
In our scenario, this will be the camera flash of the Android smartphone. 
If the feeds are available, the application becomes usable and can be started by clicking a button on the dashboard.
The application will then periodically retrieve the accelerometer data and compare it to the previous measurements in order to detect a ``shake'' gesture.
When the shake gesture is identified, the application will toggle the state of the \texttt{ON/OFF} feed and stops the measurements for 2 seconds. 
We have used this application to evaluate our platform with respect to feasibility of managing the data flows using the IoT Hub API.

\begin{figure}[!t]
 \centering
 \includegraphics[width=\linewidth]{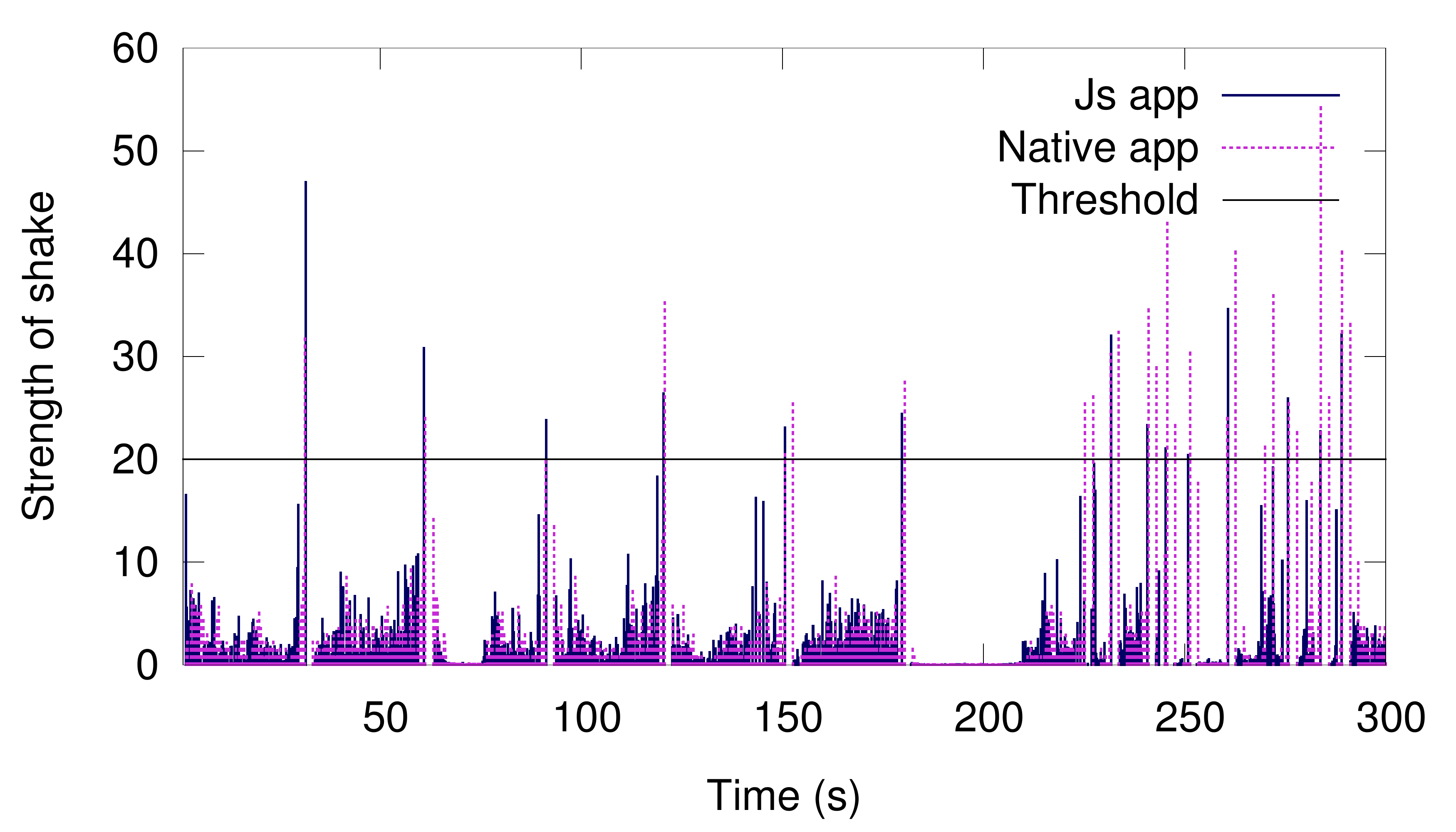}
 \caption{Comparison between native and javascript applications}
 \label{fig:evaluation}
\end{figure}

An example of the comparison between the native (developed with the Android SDK on a Samsumg Galaxy S3 mini) and the javascript applications is shown in \figurename{~\ref{fig:evaluation}}. 
In this experiment, the user shook his phone, after starting the applications (starting the javascript application automatically launches the native application), repeatedly at various frequencies.
In~\figurename{~\ref{fig:evaluation}}, the force represents the absolute difference between the aggregated $x$, $y$ and $z$ of the accelerometer of two consecutive measurements (every $200ms$). 
The experiment included three stages: (i) the phone was shook every $30s$, (ii) the phone was resting on a table and (iii) the phone was shook at high frequency.

During the experiment, the javascript application is expressing very similar behavior to the native applications. 
However, some delays can be observed during the last stage of the experiment. 
These are due to the time required to access the phone's camera (the native application did not have it implemented to avoid concurrent use of the camera). 
This resulted in the javascript application to miss some of the shake gestures observed by the native application.
The experiment was repeated 20 times and has always shown identical patterns.

Thus, this demonstrate that it is feasible to manage the data flows using the IoT Hub API.
Moreover, these applications are available for all hub-based platforms and can be distributed to a wide audience via the meta-hubs.

\makeatletter{}\section{Conclusions \& Future Work}\label{sec:conclusion}

In this paper, we presented the generic IoT Hub architecture to enable interoperability between IoT solutions. The key component of the architecture is
the IoT feed that provides loose coupling of the data and assets managed by the platform. The platform owner is able to regain the ownership of the data
and the devices functionalities by dynamically managing the IoT feeds properties via the IoT Hub API. Additionally, the IoT feeds may be exposed to 
third-parties services by publishing their description to meta-hubs. Hence, meta-hubs are the cornerstones of networks of interoperable hubs with 
added functionalities to distribute applications and services based on the IoT Hub API.
We demonstrated the feasibility of controlling data flows with the IoT Hub API as native and hub-based applications expressed similar behaviors.

We envision for future work to use an IoT-dedicated DSL for the development of complex applications such as distributed analytics on a network of hubs. The applications
will include the composition of numerous data flows, sensor data corrections, data aggregation and fusion in a distributed fashion on heterogeneous middleware systems.

\bibliographystyle{IEEEtran}
\bibliography{IEEEabrv,2015_ICC}

\end{document}